# Efficient hybrid-mode excitation in plasmonic nanoantennas by tightly focused higher-order vector beams

XIAORUN ZANG,[1,2,*] GODOFREDO BAUTISTA,[2] LÉO TURQUET,[2] TERO SETÄLÄ,[1] MARTTI KAURANEN[2] AND JARI TURUNEN[1]

[1]*Institute of Photonics, University of Eastern Finland, P.O. Box 111, FI-80101 Joensuu, Finland*
[2]*Photonics Laboratory, Physics Unit, Tampere University, P.O. Box 692, FI-33014, Tampere, Finland*
[*]*xiaorun.zang@uef.fi*

**Abstract:** Efficient optical excitation of hybridized plasmon modes in nanoantennas is vital to achieve many promising functionalities, but it can be challenging due to a field-profile mismatch between the incident light and the hybrid mode. We present a general approach for efficient hybrid-mode excitation by focusing the incident light field in the basis of cylindrically polarized vector beams of various higher-order spiral phases. Such basis vector beams are described in the higher-order polarization states and Stokes parameters (both defined locally in polar coordinates), and visualized correspondingly on the higher-order Poincaré spheres. The focal field is formulated exclusively in cylindrical coordinates as a series sum of all focused beams of the associated high-order paraxial beams. Our focal field decomposition enables an analysis of hybrid-mode excitation via higher-order vector beams, and thus yields a straightforward design of effective mode-matching field profile in the tightly focused region.

## 1. Introduction

Hybridized plasmon modes, arising from optical near-field coupling in plasmonic nanoparticle assemblies [1, 2] are attractive for a variety of applications because of their unique spectral and radiation characteristics [2–4]. Effective hybrid-mode excitation is crucial in achieving functionalities that rely on characteristics of such modes, which can be challenging as it sometimes requires a specially engineered incident light field when the conventional ones do not work. For instance, the bonding and antibonding modes are formed in nanorod dimers [2], and the former can be efficiently excited by a plane wave at normal incidence, whereas the latter's excitation is symmetry-forbidden for the same illumination but allowed by an obliquely incident plane wave [5], a localized emitter [6], or a tightly focused pulsed laser beam [7]. Another example is the use of cylindrical vector beams for the excitation of dark modes in nanorod trimers [8] or in other plasmonic clusters [9]. In general, to efficiently excite a hybrid mode, it is essential to shape the field profile of the incident light so that it effectively matches the hybrid mode's profile, which can have a significant space-varying polarization distribution over the sub-wavelength range around the nanoparticles.

Vector beams (VBs) [10, 11] with tuned spatial distributions of the polarization state add new ingredients into the light-matter interaction. Recently, they have found a considerable amount of applications in diverse research fields, including optical microscopy [12–14], optical communication [15–18], trapping [19, 20], surface plasmon polaritons [**?**, 21–23] and nonlinear nano-optics [24–28]. Typical examples are the radially and azimuthally polarized VBs, where the electric field vectors point into the radial and azimuthal directions in the beam cross section, respectively [10]. They are usually regarded as two fundamental zeroth-order instances of more general higher-order cylindrical VBs that can be geometrically illustrated on the higher-order Poincaré (HOP) spheres [29]. Compared with the standard Poincaré sphere, every point on the HOP sphere corresponds to a higher-order VB which can possess an extra helical phase modulation in addition to the spatially inhomogeneous polarization distribution. General cylindrical VBs with arbitrary higher-order polarization states have been recently extensively investigated, as they have become experimentally accessible [30–34].

Here, we investigate the shaping of the incident light with an effective mode-matching field profile through a decomposition into higher-order, cylindrically polarized VBs. We adopt the concepts of higher-order polarization states, Stokes parameters, and HOP spheres in polar coordinates on the basis of higher-order radial and azimuthal polarization states. Regarding this adoption, each individual higher-order basis VB has a locally identical polarization with respect to polar coordinates. Moreover, the same points on the adopted local HOP spheres of different higher-orders represent a group of higher-order VBs of locally identical polarization but distinct helical phase distributions. This is in contrast to the recently introduced HOP sphere [29, 34] which is defined with respect to globally orthonormal circular polarization states, where polarization distributions of the same order are generally not cylindrically symmetric and polarization distributions associated with the same points on HOP spheres of different orders are not the same either. Our higher-order VBs thus form a simple and natural basis for effectively engineering the desired field profile and, in turn, this VB decomposition enables an analysis of hybrid-mode excitation via VBs of various higher-orders polarization states.

In addition to the match of polarization and phase distributions, an efficient concentration of the input power on the sub-wavelength plasmonic nanoparticles is also important from a practical point of view. In this regard, the constructed incident beam needs to be tightly focused and the corresponding focal field is formulated exclusively in cylindrical coordinates as a series sum of focused VBs of the associated high-order cylindrical VBs on the local HOP spheres. By doing so, we trace clearly the transfer of each polarization component and the evolution of each higher-order polarization state on the local HOP sphere during tight focusing. More importantly, such a focal field formulation provides an analysis tool of the focal field profile in higher-order polarization states and it permits a forward design of mode-matching field profile with the input power effectively focused on nanoantennas. We then demonstrate that using VBs of radial polarization states of various orders is necessary to match the polarization distributions of different hybrid modes for our example of a radial plasmonic tetramer, some of which are barely coupled with radially polarized VBs of the fundamental order used in previous works [8, 9, 27]. The interaction between the tightly focused VBs and the tetramer is efficiently simulated by the boundary element method, because the focal fields are evaluated merely on the nanoparticles' surfaces.

## 2. Vector beams on local HOP spheres

We consider a monochromatic VB with harmonic time dependence $e^{-i\omega t}$, where $\omega$ is the angular frequency. The field is assumed to propagate along the $z$-axis and its (transverse) complex electric field amplitude at the waist plane can be written in polar coordinates $(r, \phi)$ as

$$\mathbf{E}_{\text{inc}}(r, \phi) = E_r(r, \phi)\hat{r} + E_\phi(r, \phi)\hat{\phi}. \tag{1}$$

Above $E_r(r, \phi)$ and $E_\phi(r, \phi)$ are the radial and azimuthal field amplitudes with $\hat{r}$ and $\hat{\phi}$ denoting the radial and azimuthal unit vectors, respectively. We assume that the amplitudes can be decomposed as $E_q(r, \phi) = E_b(r)P_q(r, \phi)$ with $q \in \{r, \phi\}$, where $E_b(r)$ is a common beam profile shared by both field components and $P_q(r, \phi)$ is a (complex) pupil function for phase and/or amplitude modulations of the corresponding field component. Upon choosing specifically the pupil functions, a VB with certain spatially varying polarization state can be engineered.

Each pupil function has an azimuthal expansion [35]

$$P_q(r,\phi) = \sum_{n=-\infty}^{+\infty} c_q^{(n)}(r) e^{in\phi}, \qquad (2)$$

where the order $n$ is an integer and the $r$-dependent expansion coefficient is

$$c_q^{(n)}(r) = \frac{1}{2\pi} \int_{-\pi}^{+\pi} P_q(r,\phi) e^{-in\phi} d\phi. \qquad (3)$$

It is convenient to introduce specific nonuniform or spatially varying polarization-state distributions for the description of VBs. Particularly relevant ones are the unit amplitude, radially and azimuthally polarized field distributions $\boldsymbol{\psi}_r$ and $\boldsymbol{\psi}_\phi$ obtained from Eq. (1) with $E_r = 1, E_\phi = 0$ and $E_r = 0, E_\phi = 1$, respectively. The basis functions $\boldsymbol{\psi}_r$ and $\boldsymbol{\psi}_\phi$ represent polarization states that are locally orthonormal [36, 37]. They enable us to construct, for example, the locally orthonormal right-hand circular polarization state $\boldsymbol{\psi}_+$ and left-hand circular counterpart $\boldsymbol{\psi}_-$ via $\boldsymbol{\psi}_\pm = (\boldsymbol{\psi}_r \pm i\boldsymbol{\psi}_\phi)/\sqrt{2}$. We further introduce the higher-order basis polarization states of unit amplitude as

$$\boldsymbol{\psi}_r^{(n)} = e^{in\phi} \boldsymbol{\psi}_r \quad \text{and} \quad \boldsymbol{\psi}_\phi^{(n)} = e^{in\phi} \boldsymbol{\psi}_\phi, \qquad (4)$$

which possess an extra spiral phase shift $n\phi$, as compared to their fundamental zeroth-order counterparts. Their linear combination forms a higher-order polarization state of the general form

$$\boldsymbol{\psi}^{(n)} = a_r^{(n)} \boldsymbol{\psi}_r^{(n)} + a_\phi^{(n)} \boldsymbol{\psi}_\phi^{(n)}, \qquad (5)$$

where $a_r^{(n)}$ and $a_\phi^{(n)}$ are the complex amplitudes associated with the corresponding $n$th-order, radial and azimuthal basis states, respectively, which can be radius-dependent. In terms of higher-order polarization states defined in Eqs. (4) and (5), the VB in Eq. (1) can be rewritten as

$$\mathbf{E}_{\text{inc}}(r,\phi) = \sum_{n=-\infty}^{+\infty} \left[ c_r^{(n)}(r) \boldsymbol{\psi}_r^{(n)} + c_\phi^{(n)}(r) \boldsymbol{\psi}_\phi^{(n)} \right] E_b(r) = \sum_{n=-\infty}^{+\infty} \boldsymbol{\psi}^{(n)} E_b(r), \qquad (6)$$

with the amplitudes $a_r^{(n)} = c_r^{(n)}(r)$ and $a_\phi^{(n)} = c_\phi^{(n)}(r)$ being functions of the radius. Such radius-dependent amplitudes permit the construction of distinctive polarization states for each individual higher-order $n$ at different $r$-values across the entire beam cross section, and summing up all higher-order VBs yields the desired field profile. On the other hand, polarization states at the same $r$-value are always locally identical, i.e., the ratio between the radial and azimuthal field amplitudes is fixed and thus the polarization distributions on an annulus are the same with respect to polar coordinates. If the ratio $c_r^{(n)}(r)/c_\phi^{(n)}(r)$ is independent of the radius, in particular, the vector fields across the entire transverse plane share the same higher-order polarization state at the order $n$.

For the polarization state of a fixed order $n$, we can introduce the Stokes parameters with respect to the higher-order basis states $\boldsymbol{\psi}_r^{(n)}$ and $\boldsymbol{\psi}_\phi^{(n)}$ as

$$S_0^{(n)} = |a_r^{(n)}|^2 + |a_\phi^{(n)}|^2, \qquad (7)$$

$$S_1^{(n)} = |a_r^{(n)}|^2 - |a_\phi^{(n)}|^2, \qquad (8)$$

$$S_2^{(n)} = 2\text{Re}\{[a_r^{(n)}]^* a_\phi^{(n)}\}, \qquad (9)$$

$$S_3^{(n)} = 2\text{Im}\{[a_r^{(n)}]^* a_\phi^{(n)}\}, \qquad (10)$$

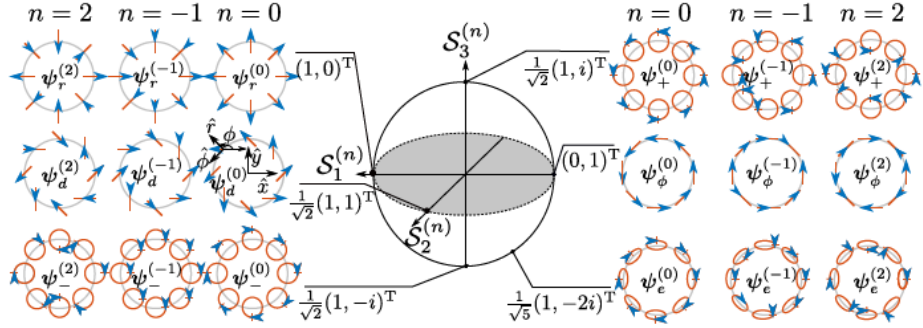

Fig. 1. The local HOP sphere representation for VBs with local polarization states of orders $n = 0, -1, 2$ (from inner to outer columns). All linear polarization states are located on the periphery of the $\mathcal{S}_1^{(n)}\mathcal{S}_2^{(n)}$-plane. The points $+\mathcal{S}_1^{(n)}$ and $-\mathcal{S}_1^{(n)}$ represent the $n$th-order, radial and azimuthal states $\psi_r^{(n)}$ and $\psi_\phi^{(n)}$, respectively. The point $+\mathcal{S}_2^{(n)}$ marks a linear state $\psi_d^{(n)}$ with equal radial and azimuthal field components. The north pole ($+\mathcal{S}_3^{(n)}$) denotes the locally right-hand circular state $\psi_+^{(n)}$ and south pole ($-\mathcal{S}_3^{(n)}$) for its left-handed counterpart $\psi_-^{(n)}$. The rest of the points are occupied by elliptical states $\psi_e^{(n)}$. The instantaneous fields are indicated by blue arrow heads. From the temporal viewpoint, the initial phases of the states with $n = -1$ advance by an amount of $\phi$ phase with respect to the corresponding states of $n = 0$, whereas the states of $n = 2$ lag behind by $2\phi$ phase.

where Re and Im are the real part and imaginary part, respectively, and the asterisk denotes the complex conjugate. The Stokes parameters in Eqs. (8)–(10) can be used for a geometric illustration of the $n$th-order polarization state $\psi^{(n)}$ that represents a beam with spatially varying but cylindrically symmetric polarization distribution, on an adopted Poincaré sphere of radius $\mathcal{S}_0^{(n)}$ (as visualized in Fig. 1). This is in contrast to the traditional Poincaré sphere representation which holds for uniformly polarized beams, and it is called the local HOP sphere in this work.

We remark that the basis polarization states of various orders defined in Eq. (4) only differ in their instantaneous field amplitudes and, as a result, identical sets of Stokes parameters of various orders represent a group of states that have the same spatial polarization distribution with respect to the local polar coordinates but distinct (delayed) temporal behaviors. As it can be seen in Fig. 1, polarization states at the same points on the local HOP spheres of various orders are represented by polarization ellipses with locally identical orientation and shape (shown as red lines, circles, or ellipses with arrows). From a temporal point of view, on the other hand, the initial phases of $n$th-order states $\psi^{(n)}$ lag (for $n > 0$) or advance (for $n < 0$) by an amount of $|n\phi|$ phase with respect to the corresponding 0th-order states $\psi^{(0)}$, as the instantaneous fields are indicated by the relative positions of the blue arrow heads on the polarization ellipses (states of the same order are shown in the same column in Fig. 1). Here, the higher-order polarization states, Stokes parameters, and local HOP sphere provide an alternative approach for describing cylindrically symmetric VBs with great simplicity and clarity by using locally identical polarization distributions, as compared with the recently introduced HOP sphere [29, 34] in which the higher-order polarization states are defined with respect to globally orthonormal circular polarization states.

## 3. Tight focusing of vector beams

We consider the tight focusing of VBs by a high numerical aperture (NA) aplanatic lens of focal length $f$ shown in Fig. 2(a). The objective is represented by a reference sphere shown as a light blue spherical cap. The focal field at an arbitrary point $\mathbf{R}_f$ near the focus in cylindrical coordinates $(\rho, \varphi, z)$ can be obtained from the Richards–Wolf formalism [38–40] by integrating the reference field $\mathbf{E}_\infty(\phi, \theta)$ over the aperture,

$$\mathbf{E}_f(\rho, \varphi, z) = -\frac{ik}{2\pi} \int_0^{\theta_{max}} \int_0^{2\pi} \mathbf{E}_\infty e^{i\mathbf{k}\cdot\mathbf{R}_f} \sin\theta \, d\phi \, d\theta, \tag{11}$$

where $\mathbf{k}$ is the wave vector, $\mathbf{k}\cdot\mathbf{R}_f = -k\rho\sin\theta\cos(\phi-\varphi) + kz\cos\theta$ (see Fig. 2), and wave number is $k = \sqrt{\mathbf{k}\cdot\mathbf{k}} = n_f k_0$ with $n_f$ being the refractive index after the reference sphere and $k_0$ the vacuum wavenumber. The maximum angle subtended by the aperture is $\theta_{max} = \arcsin(\mathrm{NA}/n_f)$ with $\mathrm{NA} = n_f \sin\theta_{max}$, and we assume the beam fills the whole aperture.

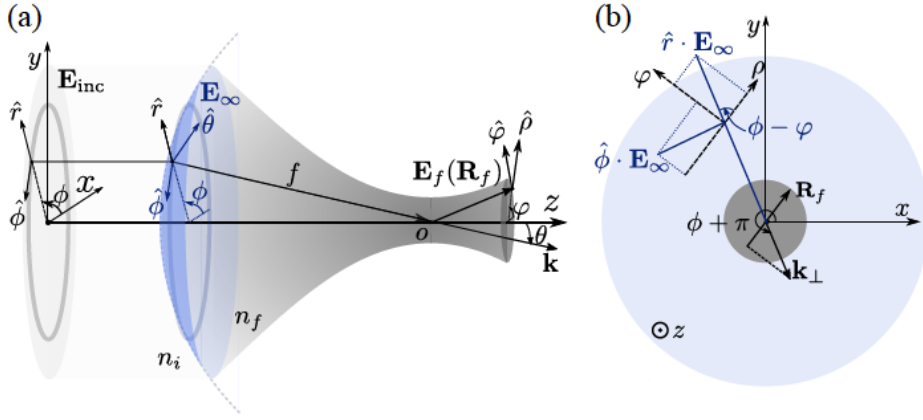

Fig. 2. Tight focusing of a vector beam in an aplanatic system. (a) The incident beam $\mathbf{E}_{inc}$ is mapped to the reference field $\mathbf{E}_\infty$ (on the light blue spherical cap with its radius being equal to the focal length $f$), the unit vectors $\hat{r}$ and $\hat{\theta}$ are in the same meridional plane, and $\hat{\phi}$ and $\hat{\theta}$ are tangential to the reference sphere. The refractive indices before and after the reference sphere are $n_i$ and $n_f$, respectively. The focal field $\mathbf{E}_f$ is sought at a point $\mathbf{R}_f$ near the focus, where $(\rho, \varphi, z)$ form the cylindrical coordinates and $\hat{\rho}$ and $\hat{\varphi}$ are the associated unit vectors. (b) The field components $\hat{r}\cdot\mathbf{E}_\infty$ and $\hat{\phi}\cdot\mathbf{E}_\infty$ are projected into the cylindrical coordinates $(\rho, \varphi, z)$. The unit vectors $\hat{r}$ and $\hat{\rho}$ make an angle $\phi - \varphi$. The wave vector $\mathbf{k}$ measures an angle $\theta$ with respect to the $z$-axis and its transverse projection $\mathbf{k}_\perp$ makes a $\phi + \pi$ angle to the $x$-axis.

In refraction at the aplanatic lens, following Richards and Wolf [39], the reference field amplitude vector writes as $\mathbf{E}_\infty(\theta, \phi) = f\sqrt{n_i/n_f}\sqrt{\cos\theta}\,(E_r\hat{\theta} + E_\phi\hat{\phi})$ where $n_i$ is the refractive index before the reference sphere, and the sine condition and energy conservation are used. In the column vector representation in the basis $[\hat{r}, \hat{\phi}, \hat{z}]$ the reference field reads as

$$\begin{bmatrix} \hat{r}\cdot\mathbf{E}_\infty(\theta, \phi) \\ \hat{\phi}\cdot\mathbf{E}_\infty(\theta, \phi) \\ \hat{z}\cdot\mathbf{E}_\infty(\theta, \phi) \end{bmatrix} = \mathbf{L}(\theta) \begin{bmatrix} E_r(f\sin\theta, \phi) \\ E_\phi(f\sin\theta, \phi) \end{bmatrix}, \tag{12}$$

where the relations $r = f \sin\theta$ and $\hat{\theta} = \hat{r}\cos\theta + \hat{z}\sin\theta$ are used, and

$$\mathbf{L}(\theta) = \begin{bmatrix} L_{rr}(\theta) & L_{r\phi}(\theta) \\ L_{\phi r}(\theta) & L_{\phi\phi}(\theta) \\ L_{zr}(\theta) & L_{z\phi}(\theta) \end{bmatrix} = f\sqrt{\frac{n_i}{n_f}}\sqrt{\cos\theta}\begin{bmatrix} \cos\theta & 0 \\ 0 & 1 \\ \sin\theta & 0 \end{bmatrix} \tag{13}$$

can be understood as the matrix representation for the refraction at the aplanatic lens. The matrix element $L_{sq}(\theta)$ with $s \in \{r, \phi, z\}$ and $q \in \{r, \phi\}$ denotes the field amplitude transfer from the $q$-component of the incident VB to the $s$-component of the reference field.

To perform the integration in Eq. (11), the reference field needs to be expanded in the same basis as the focal field. A convenient choice is the basis $[\hat{\rho}, \hat{\varphi}, \hat{z}]$, in which the column vector form of Eq. (11) becomes

$$\begin{bmatrix} \hat{\rho} \cdot \mathbf{E}_f(\rho, \varphi, z) \\ \hat{\varphi} \cdot \mathbf{E}_f(\rho, \varphi, z) \\ \hat{z} \cdot \mathbf{E}_f(\rho, \varphi, z) \end{bmatrix} = -\frac{ik}{2\pi} \int_0^{\theta_{max}} \int_0^{2\pi} \begin{bmatrix} \hat{\rho} \cdot \mathbf{E}_\infty(\theta, \phi) \\ \hat{\varphi} \cdot \mathbf{E}_\infty(\theta, \phi) \\ \hat{z} \cdot \mathbf{E}_\infty(\theta, \phi) \end{bmatrix} e^{i\mathbf{k}\cdot\mathbf{R}_f} \sin\theta d\phi d\theta. \tag{14}$$

Above the vector form of $\mathbf{E}_\infty(\theta, \phi)$ in the basis $[\hat{\rho}, \hat{\varphi}, \hat{z}]$ is connected to its representation in the basis $[\hat{r}, \hat{\phi}, \hat{z}]$ through

$$\begin{bmatrix} \hat{\rho} \cdot \mathbf{E}_\infty(\theta, \phi) \\ \hat{\varphi} \cdot \mathbf{E}_\infty(\theta, \phi) \\ \hat{z} \cdot \mathbf{E}_\infty(\theta, \phi) \end{bmatrix} = \mathbf{T}(\phi, \varphi) \begin{bmatrix} \hat{r} \cdot \mathbf{E}_\infty(\theta, \phi) \\ \hat{\phi} \cdot \mathbf{E}_\infty(\theta, \phi) \\ \hat{z} \cdot \mathbf{E}_\infty(\theta, \phi) \end{bmatrix} \tag{15}$$

with the connection matrix

$$\mathbf{T}(\phi, \varphi) = \begin{bmatrix} T_{\rho r} & T_{\rho\phi} & T_{\rho z} \\ T_{\varphi r} & T_{\varphi\phi} & T_{\varphi z} \\ T_{zr} & T_{z\phi} & T_{zz} \end{bmatrix} = \begin{bmatrix} \cos(\phi-\varphi) & -\sin(\phi-\varphi) & 0 \\ \sin(\phi-\varphi) & \cos(\phi-\varphi) & 0 \\ 0 & 0 & 1 \end{bmatrix}, \tag{16}$$

obtained by recognizing, from Fig. 2, the relations of field components

$$\hat{\rho} \cdot \mathbf{E}_\infty = \hat{r} \cdot \mathbf{E}_\infty \cos(\phi-\varphi) - \hat{\phi} \cdot \mathbf{E}_\infty \sin(\phi-\varphi) \tag{17}$$

and

$$\hat{\varphi} \cdot \mathbf{E}_\infty = \hat{r} \cdot \mathbf{E}_\infty \sin(\phi-\varphi) + \hat{\phi} \cdot \mathbf{E}_\infty \cos(\phi-\varphi). \tag{18}$$

The matrix element $T_{ps}(\phi, \varphi)$ with $p \in \{\rho, \varphi, z\}$ and $s \in \{r, \phi, z\}$ characterizes the focal field's $p$-component that arises from the $s$-component of the reference field.

Substituting Eq. (12) into Eq. (15), as well as using Eqs. (1) and (6), the focal field in vector form in the basis $[\hat{\rho}, \hat{\varphi}, \hat{z}]$ writes as

$$\begin{bmatrix} \hat{\rho} \cdot \mathbf{E}_f(\rho, \varphi, z) \\ \hat{\varphi} \cdot \mathbf{E}_f(\rho, \varphi, z) \\ \hat{z} \cdot \mathbf{E}_f(\rho, \varphi, z) \end{bmatrix} = -\frac{ik}{2\pi} \int_0^{\theta_{max}} \int_0^{2\pi} \mathbf{T}(\phi, \varphi)\mathbf{L}(\theta) \sum_{n=-\infty}^{+\infty} \begin{bmatrix} c_r^{(n)}(f\sin\theta) \\ c_\phi^{(n)}(f\sin\theta) \end{bmatrix} e^{in\phi}$$
$$\times E_b(f\sin\theta) e^{-ik\rho\sin\theta\cos(\phi-\varphi)} e^{ikz\cos\theta} \sin\theta d\phi d\theta. \tag{19}$$

Performing the integration of the $\phi$-dependent portion at a given order $n$, we can write

$$\frac{-\mathrm{i}k}{2\pi}\int_0^{2\pi}\mathbf{T}(\phi,\varphi)e^{\mathrm{i}n\phi}e^{-\mathrm{i}k\rho\sin\theta\cos(\phi-\varphi)}\mathrm{d}\phi = e^{\mathrm{i}n\varphi}\frac{(-\mathrm{i})^n k}{2}\begin{bmatrix}(J_{n-1}-J_{n+1}) & -\mathrm{i}(J_{n-1}+J_{n+1}) & 0 \\ \mathrm{i}(J_{n-1}+J_{n+1}) & (J_{n-1}-J_{n+1}) & 0 \\ 0 & 0 & 2J_n\end{bmatrix}$$

$$= e^{\mathrm{i}n\varphi}\begin{bmatrix}\Theta^{(n)}_{\rho r} & \Theta^{(n)}_{\rho\phi} & \Theta^{(n)}_{\rho z} \\ \Theta^{(n)}_{\varphi r} & \Theta^{(n)}_{\varphi\phi} & \Theta^{(n)}_{\varphi z} \\ \Theta^{(n)}_{z r} & \Theta^{(n)}_{z\phi} & \Theta^{(n)}_{zz}\end{bmatrix} = e^{\mathrm{i}n\varphi}\mathbf{\Theta}^{(n)}, \qquad (20)$$

where $J_n = J_n(k\rho\sin\theta)$ denotes the $n$th order Bessel function with argument $k\rho\sin\theta$ for brevity, the $n$th-order matrix $\mathbf{\Theta}^{(n)}(\theta,\rho,\varphi)$ and its elements are introduced such that the above equation holds, and we have used the integral representation of the Bessel function

$$\int_0^{2\pi}e^{\mathrm{i}n\alpha}e^{-\mathrm{i}x\cos\alpha}\mathrm{d}\alpha = \int_0^{2\pi}e^{\mathrm{i}n\alpha}e^{\mathrm{i}x\cos(\alpha+\pi)}\mathrm{d}\alpha = 2\pi(-\mathrm{i})^n J_n(x). \qquad (21)$$

Expressing the remaining integration over $\theta$ for a given order $n$ compactly, we introduce

$$I^{(n)}_p(\rho,z) = \int_0^{\theta_{\max}}\sum_{s,q}\Theta^{(n)}_{ps}(\theta,\rho,\varphi)L_{sq}(\theta)c^{(n)}_q(f\sin\theta)E_b(f\sin\theta)e^{\mathrm{i}kz\cos\theta}\sin\theta\mathrm{d}\theta, \qquad (22)$$

with the subscripts $p \in \{\rho,\varphi,z\}$, $s \in \{r,\phi,z\}$, and $q \in \{r,\phi\}$. Overall, the focal field becomes,

$$\mathbf{E}_f(\rho,\varphi,z) = \sum_n \mathbf{E}^{(n)}_f = \sum_n [I^{(n)}_\rho(\rho,z)\boldsymbol{\psi}^{(n)}_\rho + I^{(n)}_\varphi(\rho,z)\boldsymbol{\psi}^{(n)}_\varphi + I^{(n)}_z(\rho,z)e^{\mathrm{i}n\varphi}\hat{z}], \qquad (23)$$

where the $n$th-order transverse field is expanded on the higher-order basis states $\boldsymbol{\psi}^{(n)}_\rho = e^{\mathrm{i}n\varphi}\boldsymbol{\psi}_\rho$ and $\boldsymbol{\psi}^{(n)}_\varphi = e^{\mathrm{i}n\varphi}\boldsymbol{\psi}_\varphi$ with $\boldsymbol{\psi}_\rho$ and $\boldsymbol{\psi}_\varphi$ being the unit-amplitude, radially and azimuthally polarized field distributions with respect to the cylindrical coordinates $(\rho,\varphi,z)$. The tightly focused field also acquires a longitudinal component that has the same helical phase variation as the transverse focal field. Compared with the fundamental order longitudinal field (so-called optical needle [26, 41]), the higher-order ones may have profound impacts in applications such as second-harmonic generation [24, 42] and single emitter probing in microscopy [13].

For hybrid-mode excitation of in-plane nanoantennas, we can consider only the transverse focal field whose spatially varying polarization distributions can be studied with the Stokes parameters introduced in Eqs. (7)–(10), where the associated complex amplitudes are $a^{(n)}_\rho = I^{(n)}_\rho(\rho,z)$ and $a^{(n)}_\varphi = I^{(n)}_\varphi(\rho,z)$. In addition, the incident field at the $n$th-order polarization state [in Eq. (6)] is converted to the focal field at the polarization state of the same order $n$ [in Eq. (23)]. This implies that the incident VB and the transverse part of the focal field have the same amount of helical phase or, in other words, the same orbital angular momentum. But, the transverse focal field generally does not preserve the same local polarization distributions as the incident VB. This results from energy exchanges between the radial and azimuthal field components in the focusing process due to the presence of nonzero elements $\Theta^{(n)}_{\varphi r}$ and $\Theta^{(n)}_{\rho\phi}$ for $n \neq 0$. Furthermore, the polarization distributions are both $\rho$- and $z$-dependent according to Eq. (22). Consequently, the transverse field near the focus generally does not have a pure radial or azimuthal polarization distribution except for the fundamental zeroth-order.

A typical example of paraxial VBs is a cylindrically polarized Laguerre-Gaussian (CPLG) beam [43], where the common beam profile $E_b(r) = \exp^{-\tau^2/2}$ is a Gaussian with $\tau = \sqrt{2}r/w$ and $w$ being the radius of the beam waist. For this beam, the $n$th-order expansion coefficient in Eq. (2) takes the following explicit form:

$$c_q^{(n)}(r) = A_q^{(n)} \tau^{n\mp 1} L_m^{n\mp 1}(\tau^2), \tag{24}$$

with $A_q^{(n)}$ being the amplitude of the corresponding field component, $L_m^{n\mp 1}$ the generalized Laguerre polynomials, $m$ the radial index, and $n \mp 1$ the azimuthal index. For the fundamental order when $n = 0$, the plus sign should be used in the power and azimuthal order so that $c_q^{(0)}(r) \propto \tau L_m^1(\tau^2)$ yields a null field on the optical axis where the phase has a singularity. We use the explicit form of the expansion coefficients in Eq. (24) and choose the "+" sign for all simulations in this paper.

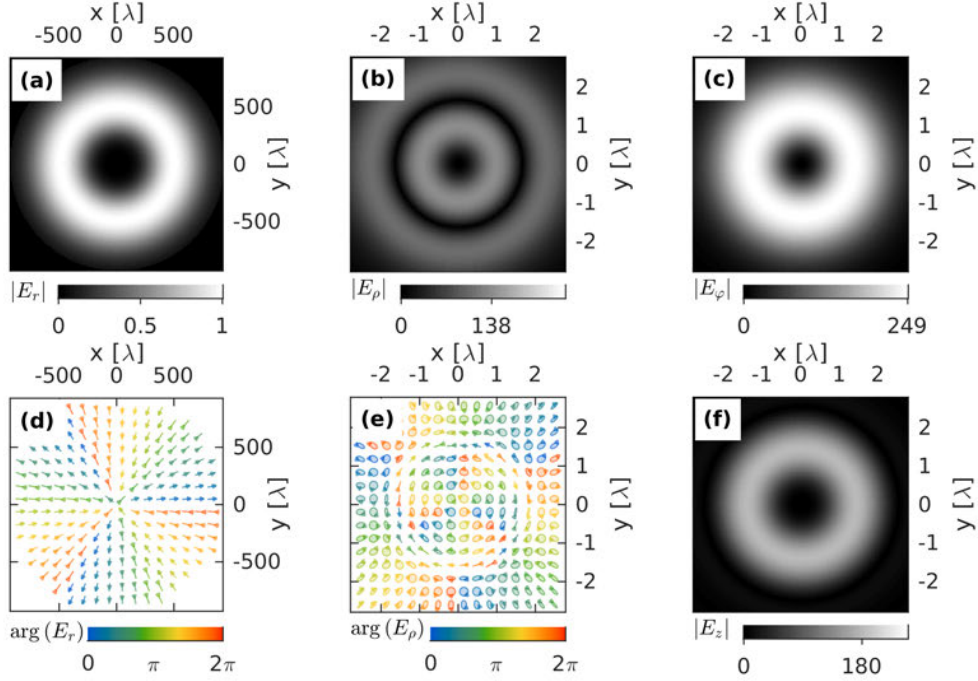

Fig. 3. Amplitude and polarization distributions of $\psi_r^{(3)}$ and its focused VBs. Field amplitudes are shown in greyscale colormaps for (a) the incident radial $|E_r|$, (b) the focused radial $|E_\rho|$, (c) the focused azimuthal $|E_\varphi|$, and (f) the focused longitudinal $|E_z|$ components, respectively. The polarization distributions are shown in (d) for the paraxial and (e) the focused transverse fields, where the phases of the radial components are indicated in blue to red colormaps. The blue or red ones imply locally in-phase instantaneous fields and the light green ones indicate out-of-phase cases.

Figure 3 illustrates the field amplitude and polarization distributions of a $\psi_r^{(3)}$ beam and its focused counterpart. In detail, the incident VB has a waist radius of $w = 0.4$ mm at the wavelength $\lambda = 1060$ nm, the radial index is $m = 0$, the azimuthal index is $n + 1 = 4$, and the azimuthal field amplitude $A_\phi^{(3)}$ is set to zero. In addition, the objective has an NA of 0.8 and the focal length is $f = 1$ mm. The radial field amplitude $|E_r|$ of the $\psi_r^{(3)}$ beam is shown in greyscale in Fig. 3(a), and its polarization distribution is shown in Fig. 3(d) on which the phase distribution of the radial field $\arg(E_r)$ is also superimposed in a blue to red colormap. Radial and azimuthal field amplitudes $|E_\rho|$ and $|E_\varphi|$ in the focal plane generated in tight focusing are shown in Fig. 3(b) and (c), respectively. As indicated by the greyscale colormaps, the azimuthal and radial field amplitudes depend differently on the radius and, for some radii, the azimuthal amplitudes even exceed its radial counterparts. Consequently, the transverse focal field is in general polarization state as shown in Fig. 3(e). Nevertheless, the order of polarization states remains unchanged throughout the focusing process. In other words, the field in Fig. 3(d) shares the same higher-order radial polarization state that is represented at the point $+\mathcal{S}_1^{(3)}$ on the local HOP sphere. By contrast, the polarization states of the transverse focal fields in Fig. 3(e) spread over the local HOP sphere of the same order. As the radius $\rho$ increases, the point on the local HOP sphere representing the polarization state of the transverse focal field shifts along the meridian from the north pole $+\mathcal{S}_3^{(3)}$ (the third-order right-hand circular polarization) to the south pole $-\mathcal{S}_3^{(3)}$ (the third-order left-hand circular polarization) via the point $-\mathcal{S}_1^{(3)}$ (the third-order azimuthal polarization), and then back to the north pole $+\mathcal{S}_3^{(3)}$ via the point $+\mathcal{S}_1^{(3)}$ (the third-order radial polarization). At last, a significant longitudinal field in the focal region is also shown in Fig. 3(f). Such a strong longitudinal component twists the major axes of the polarization ellipses out of the focal plane onto a Möbius strip [44] at a given $r$-value.

## 4. Efficient hybrid mode excitation from plasmonic nanoantennas

In this section, we proceed to show that focused VBs can efficiently excite distinct hybrid modes in plasmonic nanoantennas by providing not only matching polarization distributions but also strong incident field on the nanoparticles. Compared with the previous works that used cylindrical VBs (of the fundamental order when analyzed in terms of the higher-order polarization states) [8, 9, 27], here we demonstrate the importance of using VBs in higher-order polarization states to interact with a number of hybrid modes. As an example, we theoretically study light scattering from a plasmonic tetramer consisting of four identical, radially oriented gold nanorods. Each individual nanorod supports a dipole-like localized surface plasmon mode that is associated with the rod's long axis and can be excited by a matching, linearly polarized plane wave or Gaussian beam. In the radial tetramer, a number of hybridized modes of cylindrically symmetric polarization distributions emerge from near-field coupling among the dipolar modes. We compare the excitation efficiencies of various hybrid modes in the tetramer when illuminated by paraxial VBs on the fundamental zeroth-order, first-order, and second-order radial polarization states, as well as the corresponding focused beams. The excitation efficiencies are characterized by the total powers scattered from the tetramer, which are numerically calculated by the boundary element method [45, 46]. We use the formulations given in the previous sections to calculate the focal electric fields. The associated magnetic fields in the focal region are readily obtained by replacing $\mathbf{E}_f$ with $\mathbf{H}_f$ and $\mathbf{E}_\infty$ with $\mathbf{H}_\infty = (1/Z_f)(\mathbf{k}/k) \times \mathbf{E}_\infty$ in Eq. (11) [38], where $Z_f$ is the wave impedance after the reference sphere. From a computational point of view, the boundary element method is very efficient for modeling the interaction between tightly focused beams and plasmonic nanoantennas, since the excitation focal fields are evaluated only on the surfaces of the nanoparticles. In detail, only the electric and magnetic fields in the embedding medium at nodes defined in the Gaussian quadrature rules are calculated for the evaluation of the surface integral in Eqs. (5.27) and (5.28) of the reference [46]. In all simulations, the same parameters as in Fig. 3 are used. In addition, the middle plane of the tetramer coincides with the focal plane where $z = 0$. Each constituent nanorod has a width of 35 nm, thickness of 20 nm, and length of 165 nm. Each pair of opposite nanorods has an end-to-end separation of 35 nm, and the smallest distance of every neighboring nanorods are then 17.5 nm between rounded corners. The nanorods are embedded in a medium of effective refractive index $n_f = 1.26$, and the refractive index of gold nanorods is taken from the tabulated data [47].

The radial field amplitude distributions of the paraxial $\boldsymbol{\psi}_r^{(0)}$, $\boldsymbol{\psi}_r^{(1)}$, and $\boldsymbol{\psi}_r^{(2)}$ VBs are shown in Fig. 4(a) for a radius range from 0 to 600 µm, as well as in Fig. 4(b) for the radius range around the nanorods with $r \in [0, 240]$ nm. Each paraxial VB has a unit peak-field amplitude and its energy is mainly distributed around $200 - 400$ µm away from the optical axis. The field amplitudes on the nanorods [as the nanorod's center is indicated by the vertical dashed line in Fig. 4(b)] become very weak; they drop down to $\sim 10^{-4}$, $10^{-7}$ and $10^{-10}$ V/m for the paraxial $\psi_r^{(0)}$, $\psi_r^{(1)}$, and $\psi_r^{(2)}$ VBs, respectively. The focused radial fields of the paraxial $\psi_r^{(0)}$, $\psi_r^{(1)}$, and $\psi_r^{(2)}$ VBs are plotted in Fig. 4(c) and (d) for the radius ranges of $[0, 2]$ µm and $[0, 240]$ nm, separately. It is seen that the focused VB's energy concentrates down to the wavelength scale, and the field amplitudes around the nanorods are $\sim 10^2$ V/m which are several orders of magnitude higher than those of paraxial counterparts.

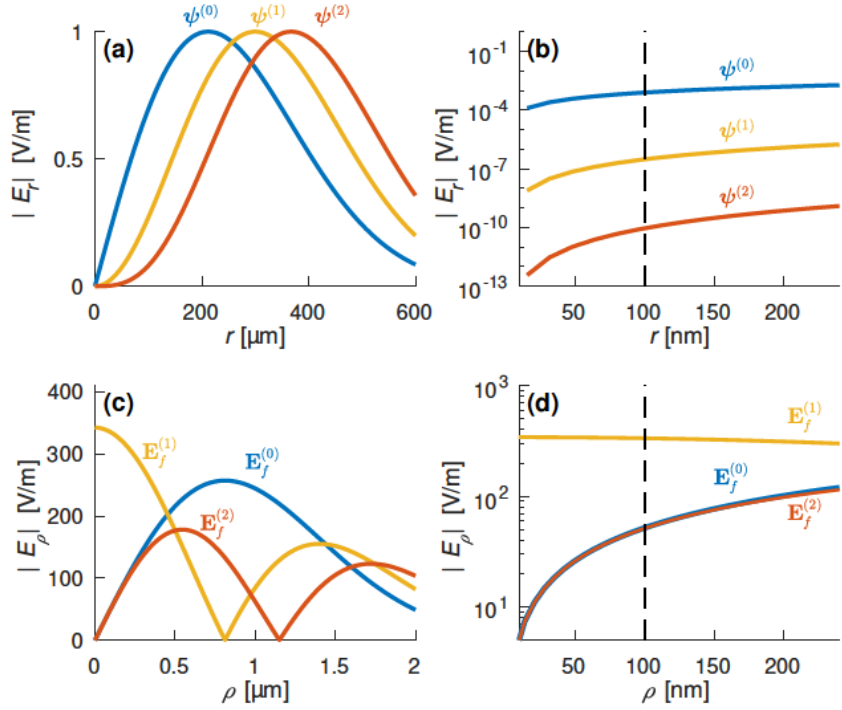

Fig. 4. The radial field amplitude distributions of the paraxial $\psi_r^{(0)}$, $\psi_r^{(1)}$, and $\psi_r^{(2)}$ VBs are shown in (a) and (b), and the focused counterparts denoted by symbols $\mathbf{E}_f^{(0)}$, $\mathbf{E}_f^{(1)}$, and $\mathbf{E}_f^{(2)}$, respectively, are displayed in (c) and (d). The vertical dashed lines in (b) and (d) mark the nanorod's center.

The polarization distributions of the paraxial $\psi_r^{(0)}$, $\psi_r^{(1)}$, and $\psi_r^{(2)}$ VBs are shown in Fig. 5(a)–(c), and the phase distributions are visualized in blue to red colormaps. The paraxial VBs are in the polarization states represented by the points $+S_1^{(0)}$, $+S_1^{(1)}$, and $+S_1^{(2)}$, respectively, on the local HOP spheres of the corresponding orders, and their electric fields are all radially polarized that match individually with the localized surface plasmons along the nanorod's long axes. The instantaneous electric fields of the fundamental $\psi_r^{(0)}$ VB on all nanorods are in phase (drawn as blue arrows). It yields an excitation of a hybrid mode that features two pairs of in-phase, antibonding localized surface plasmons with a blue-shifted resonant wavelength at $\sim 760$ nm [see the blue solid curve in Fig. 5(g)], as compared to the plasmon resonance of an independent nanorod at $\sim 860$ nm (the black dashed curve). For the first-order $\psi_r^{(1)}$ VB, the instantaneous fields on opposite nanorods are $\pi$ out-of-phase which are visualized by either blue to light green arrows or dark green to yellow arrows. The hybrid mode excited by the $\psi_r^{(1)}$ VB has a red-shifted resonance at $\sim 930$ nm [the yellow curve in Fig. 5(g)], arising from a bonding localized surface plasmon that rotates, as a function of time, between two pairs of opposite nanorods. In the case of the second-order $\psi_r^{(2)}$ VB, the neighboring nanorods experience $\pi$ out-of-phase instantaneous fields, and two pairs of out-of-phase, antibonding localized surface plasmons emerge and contribute to a hybrid mode whose resonance is red-shifted to $\sim 970$ nm [the red curve in Fig. 5(g)]. The paraxial VBs, when their polarization distributions are appropriately tuned, are able to excite various hybrid modes that possess distinct spectra with well-separated resonant wavelengths. However, the corresponding excitation efficiencies are extremely low, as indicated by the multiplication factors ($5.3 \times 10^{-11}$, $1.0 \times 10^{-17}$, and $1.4 \times 10^{-24}$ on the blue, yellow, and red spectral curves, respectively) for obtaining the absolute values of the total scattered power. The low efficiencies are attributed to the weak incident field amplitudes effectively exerted on the nanorods, regardless of the perfect matches of polarization distributions between the incident fields and the localized surface plasmons of the hybrid modes. In detail, the field amplitudes on the nanorods' centers are only $7.8 \times 10^{-4}$, $3.0 \times 10^{-7}$, and $9.2 \times 10^{-11}$ V/m [see the factors around the dashed circles in Fig. 5(a)–(c)] for the paraxial $\psi_r^{(0)}$, $\psi_r^{(1)}$, and $\psi_r^{(2)}$ VBs, separately.

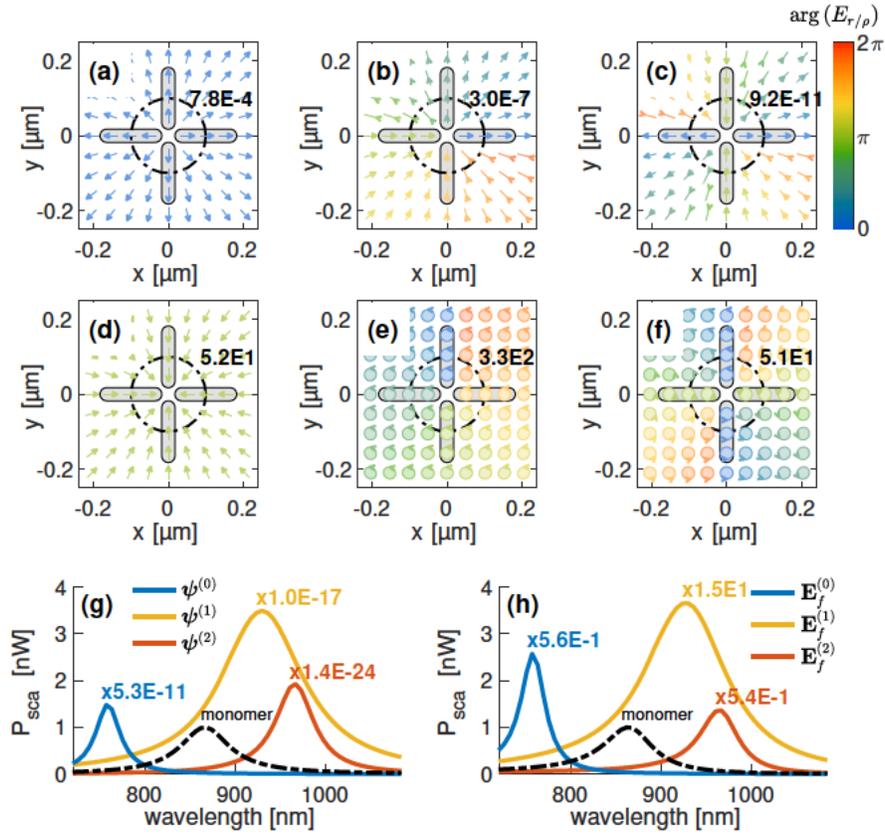

Fig. 5. The excitation of hybrid modes in the radial tetramer. The polarization distributions around the tetramer (solid black lines) in the focal plane are shown in (a)–(c) for the paraxial VBs and (d)–(f) for the focused VBs. The phases of the radial fields are visualized in blue to red colormaps. The radial field amplitudes on nanorods' centers (crossed by black dashed curves) are $7.8 \times 10^{-4}$ in (a), $3 \times 10^{-7}$ in (b), $9.2 \times 10^{-11}$ in (c), 52 in (d), 330 in (e), and 51 V/m in (f). The total powers scattered from the tetramer are plotted in the blue, yellow, and red curves in (g) when illuminated by the paraxial $\psi_r^{(0)}$, $\psi_r^{(1)}$, and $\psi_r^{(2)}$ VBs of unit-amplitude peak field or (h) the corresponding focused $\mathbf{E}_f^{(0)}$, $\mathbf{E}_f^{(1)}$, and $\mathbf{E}_f^{(2)}$ VBs, respectively. The power spectra of a single independent nanorod (monomer) is plotted as a reference in the black dashed curve. The absolute power values are obtainable when multiplied by the factors near each spectral peak.

The situation is significantly improved when the paraxial VBs are tightly focused. Figure 5(d)–(f) display the polarization distributions of the transverse fields and the phase distributions of the radial fields of the $\mathbf{E}_f^{(0)}$, $\mathbf{E}_f^{(1)}$, and $\mathbf{E}_f^{(2)}$ VBs that are tightly focused from the paraxial $\psi_r^{(0)}$, $\psi_r^{(1)}$, and $\psi_r^{(2)}$ VBs, respectively. The factors close to the dashed circles mark the radial field amplitudes on nanorods' centers, which are 52, 330, and 51 V/m for the focused $\mathbf{E}_f^{(0)}$, $\mathbf{E}_f^{(1)}$, and $\mathbf{E}_f^{(2)}$ VBs, separately. The strong, effective radial fields of the focused VBs significantly elevate the excitation efficiencies of the associated hybrid modes by approximately 10, 18, and 23 orders of magnitude, as seen from Fig. 5(g) and (h).

The focused $\mathbf{E}_f^{(0)}$ VB's transverse field is in the polarization state that is represented by the same $+\mathcal{S}_1^{(0)}$ point on the local HOP sphere as its paraxial counterpart except for field concentration and enhancement (i.e., the local HOP sphere's radius $\mathcal{S}_0^{(0)}$ increases) on the nanoscale tetramer. Therefore, it interacts with the same hybrid mode as its paraxial counterpart except for yielding significantly high efficiency. For both the focused $\mathbf{E}_f^{(1)}$ and $\mathbf{E}_f^{(2)}$ VBs, the azimuthal field components emerge from the nonzero $I_\varphi^{(1)}(\rho,0)$ and $I_\varphi^{(2)}(\rho,0)$ in Eq. (23). Furthermore, the ratios $I_\rho^{(1)}(\rho,0)/I_\varphi^{(1)}(\rho,0)$ and $I_\rho^{(2)}(\rho,0)/I_\varphi^{(2)}(\rho,0)$ are radius-dependent. Therefore, the polarization states of their transverse fields become elliptically or even circularly polarized [see the polarization ellipses in Fig. 5(e) and (f)] which can be represented by points near the north poles $+\mathcal{S}_3^{(1)}$ and $+\mathcal{S}_3^{(2)}$ on the corresponding local HOP spheres. Since the azimuthal fields barely couple to the hybrid modes whose localized surface plasmons have dominant radial polarizations, they can be ignored here. The plasmon mode along the short axis of each individual nanorod, nevertheless, can be excited by the azimuthal field component, but its resonant wavelength is below 800 nm which is outside the spectral window we are considering in this work. On the other hand, the remaining transverse fields are in radial polarization states at the $+\mathcal{S}_1^{(1)}$, and $+\mathcal{S}_1^{(2)}$ points on the local HOP spheres with much larger radii $\mathcal{S}_0^{(1)}$ and $\mathcal{S}_0^{(2)}$, i.e., enhanced field amplitudes on the tetramer, than those of the paraxial counterparts. This explains the increases in excitation efficiencies for the corresponding hybrid modes. It is also worth noticing that there is actually no phase singularity for the central field of the focused $\mathbf{E}_f^{(1)}$ VB and a nonzero field on the optical axis [see the yellow curve in Fig. 4(d)] manifests itself during the tight focusing. It also implies that we could have chosen the "−" sign in Eq. (24) for the paraxial $\psi_r^{(1)}$ VB. Even in that particular choice for the paraxial $\psi_r^{(1)}$ VB where the radial field would have a peak in the center, the hybrid-mode excitation efficiency would be still a few orders lower than in the situation of using the focused VB. Therefore, tight focusing of the paraxial VBs is needed in general for obtaining high hybrid-mode excitation efficiencies.

To further emphasize the importance of higher-order VBs in hybrid-mode excitation, we next study light scattering from a radial plasmonic dimer that is formed by removing one pair of opposite nanorods from the tetramer. The calculated scattering spectra are shown in Fig. 6, with a comparison to the tetramer's spectra. The canonical anti-bonding and bonding modes in the radial dimer are discernable from the blue-shifted and red-shifted resonant wavelengths, respectively, in the scattering spectra (see the dashed blue and yellow curves in Fig. 6). Similar modes are formed in the tetramer as indicated in solid blue and yellow curves. However, a third type of hybrid mode (see the solid red curve in Fig. 6) in the tetramer is revealed when it is excited by the focused $\mathbf{E}_f^{(2)}$ VB. This marks a significant difference with the anti-bonding mode in either the dimer or the tetramer when illuminated by the focused $\mathbf{E}_f^{(0)}$ VB, and the necessity of using higher-order VBs to match both the polarization and phase distributions in hybrid-mode excitation.

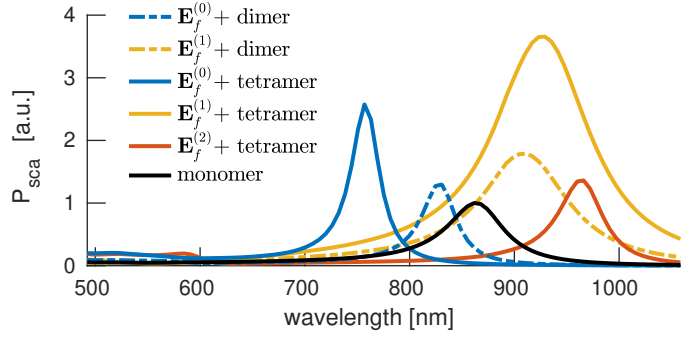

Fig. 6. Total powers scattered from the radial plasmonic dimer, tetramer, and monomer. The spectra of the dimer when illuminated by the focused $\mathbf{E}_f^{(0)}$ and $\mathbf{E}_f^{(1)}$ VBs are plotted in the dashed blue and yellow curves, respectively. The solid blue, yellow, red, and black curves represent the same spectra of the tetramer and monomer as in Fig. 5 (h).

Although we have only demonstrated the efficient excitation of hybrid modes in a radial plasmonic tetramer and dimer by using tightly focused VBs of various higher-order radial polarization states, the developed framework for shaping and analyzing the focused VB's higher-order polarization states is readily extended to cases involving hybrid modes in other plasmonic tetramers or other oligomers aggregated of nanorods or nanoparticles of other shapes. For instance, if the constituent nanoparticles are of disk shape, focused VBs of higher-order circular polarization states can be of paramount importance, because the modes in each nanodisk are symmetric that will be excited by both the radial and azimuthal field components and one would expect different hybridization in plasmon modes.

## 5. Conclusions

We investigated a general approach to engineer tightly focused VBs that can efficiently and selectively couple with various hybrid modes in plasmonic nanoantennas. We used higher-order radially and azimuthally polarized VBs as a natural basis to construct the paraxial VB of desired field profile, which was then tightly focused to effectively deliver the input power on a sub-wavelength nanoantenna where the focal field is still effective mode-matching. The desired field profile was obtained in the paraxial VB by superposing higher-order VBs and tuning the radius-dependent expansion coefficients. We adopted the concept of higher-order VBs on local HOP spheres, whose polarization distributions are cylindrically symmetric that have locally identical polarizations but distinctive instantaneous electric fields as a manifestation of the helical phase distributions. This description provides a clear understanding of the polarization transfers from the incident paraxial VBs into the focused ones, and thus the modification in the field profile was traced during focusing. Furthermore, it permits a backward optimization or inverse design to alleviate unnecessary polarizations in the focal field. We theoretically examined the efficient excitation of hybrid modes in a radial plasmonic tetramer by radially polarized paraxial VBs of various higher-orders and their focused counterparts. We explained, via our developed VB analysis in higher-order polarization states, that the corresponding focused VB shares a significant field component in the same polarization distribution as the paraxial one which permits a much higher efficiency for the associated hybrid mode. This example demonstrates the importance of higher-order VBs for the efficient, selective hybrid-mode excitation and it also indicates the great potential of tightly focused vector beams for tailoring light scattering from plasmonic nanoantennas through controlled hybrid-mode excitation.

## Acknowledgments

The authors appreciate fruitful discussions with Prof. Marco Ornigotti. This work gets the financial support from the Academy of Finland (287651) and the Flagship of Photonics Research and Innovation (PREIN) (320165, 320166).

## Disclosures

The authors declare no conflict of interest.

## References

1. M. Hentschel, M. Saliba, R. Vogelgesang, H. Giessen, A. P. Alivisatos, and N. Liu, "Transition from isolated to collective modes in plasmonic oligomers," Nano Lett. **10**, 2721–2726 (2010).
2. N. J. Halas, S. Lal, W.-S. Chang, S. Link, and P. Nordlander, "Plasmons in strongly coupled metallic nanostructures," Chem. Rev. **111**, 3913–3961 (2011).
3. C. E. Talley, J. B. Jackson, C. Oubre, N. K. Grady, C. W. Hollars, S. M. Lane, T. R. Huser, P. Nordlander, and N. J. Halas, "Surface-enhanced Raman scattering from individual Au nanoparticles and nanoparticle dimer substrates," Nano Lett. **5**, 1569–1574 (2005).
4. F. Hao, P. Nordlander, M. T. Burnett, and S. A. Maier, "Enhanced tunability and linewidth sharpening of plasmon resonances in hybridized metallic ring/disk nanocavities," Phys. Rev. B **76**, 245417 (2007).
5. W. Zhou and T. W. Odom, "Tunable subradiant lattice plasmons by out-of-plane dipolar interactions," Nat. Nanotechnol. **6**, 423–427 (2011).
6. M. Liu, T.-W. Lee, S. K. Gray, P. Guyot-Sionnest, and M. Pelton, "Excitation of dark plasmons in metal nanoparticles by a localized emitter," Phys. Rev. Lett. **102**, 107401 (2009).
7. J.-S. Huang, J. Kern, P. Geisler, P. Weinmann, M. Kamp, A. Forchel, P. Biagioni, and B. Hecht, "Mode imaging and selection in strongly coupled nanoantennas," Nano Lett. **10**, 2105–2110 (2010).
8. D. E. Gómez, Z. Q. Teo, M. Altissimo, T. J. Davis, S. Earl, and A. Roberts, "The dark side of plasmonics," Nano Lett. **13**, 3722–3728 (2013).
9. J. Sancho-Parramon and S. Bosch, "Dark modes and Fano resonances in plasmonic clusters excited by cylindrical vector beams," ACS Nano **6**, 8415–8423 (2012).
10. Q. Zhan, "Cylindrical vector beams: from mathematical concepts to applications," Adv. Opt. Photonics **1**, 1–57 (2009).
11. C. Rosales-Guzmán, B. Ndagano, and A. Forbes, "A review of complex vector light fields and their applications," J. Opt. **20**, 123001 (2018).
12. C. J. R. Sheppard and A. Choudhury, "Annular pupils, radial polarization, and superresolution," Appl. Opt. **43**, 4322–4327 (2004).
13. L. Novotny, M. R. Beversluis, K. S. Youngworth, and T. G. Brown, "Longitudinal field modes probed by single molecules," Phys. Rev. Lett. **86**, 5251–5254 (2001).
14. R. Dorn, S. Quabis, and G. Leuchs, "Sharper focus for a radially polarized light beam," Phys. Rev. Lett. **91**, 233901 (2003).
15. G. Gbur and R. K. Tyson, "Vortex beam propagation through atmospheric turbulence and topological charge conservation," J. Opt. Soc. Am. A **25**, 225–230 (2008).
16. W. Cheng, J. W. Haus, and Q. Zhan, "Propagation of vector vortex beams through a turbulent atmosphere," Opt. Express **17**, 17829–17836 (2009).
17. G. A. Tyler and R. W. Boyd, "Influence of atmospheric turbulence on the propagation of quantum states of light carrying orbital angular momentum," Opt. Lett. **34**, 142–144 (2009).
18. J. Wang, "Advances in communications using optical vortices," Photonics Res. **4**, B14–B28 (2016).
19. S. Yan and B. Yao, "Radiation forces of a highly focused radially polarized beam on spherical particles," Phys. Rev. A **76**, 053836 (2007).
20. S. Franke-Arnold, J. Leach, M. J. Padgett, V. E. Lembessis, D. Ellinas, A. J. Wright, J. M. Girkin, P. Öhberg, and A. S. Arnold, "Optical ferris wheel for ultracold atoms," Opt. Express **15**, 8619–8625 (2007).
21. Z. Man, L. Du, C. Min, Y. Zhang, C. Zhang, S. Zhu, H. P. Urbach, and X. C. Yuan, "Dynamic plasmonic beam shaping by vector beams with arbitrary locally linear polarization states," Appl. Phys. Lett. **105**, 1–6 (2014).

22. Z. Man, W. Shi, Y. Zhang, C. Zhang, C. Min, and X. C. Yuan, "Properties of surface plasmon polaritons excited by generalized cylindrical vector beams," Appl. Phys. B: Lasers Opt. **119**, 305–311 (2015).
23. Z. Man, S. Zhang, Z. Bai, Y. Zhang, X. Ge, F. Xing, Y. P. Sun, and S. Fu, "All-optical and dynamic manipulation of surface plasmon polaritons by tailoring the polarization state of incident light," Laser Phys. Lett. **16** (2019).
24. G. Bautista, M. J. Huttunen, J. Mäkitalo, J. M. Kontio, J. Simonen, and M. Kauranen, "Second-harmonic generation imaging of metal nano-objects with cylindrical vector beams," Nano Lett. **12**, 3207–3212 (2012).
25. G. Bautista and M. Kauranen, "Vector-field nonlinear microscopy of nanostructures," ACS Photonics **3**, 1351–1370 (2016).
26. L. Turquet, X. Zang, J.-P. Kakko, H. Lipsanen, G. Bautista, and M. Kauranen, "Demonstration of longitudinally polarized optical needles," Opt. Express **26**, 27572–27584 (2018).
27. G. Bautista, C. Dreser, X. Zang, D. P. Kern, M. Kauranen, and M. Fleischer, "Collective effects in second-harmonic generation from plasmonic oligomers," Nano Lett. **18**, 2571–2580 (2018).
28. R. Camacho-Morales, G. Bautista, X. Zang, L. Xu, L. Turquet, A. Miroshnichenko, H. H. Tan, A. Lamprianidis, M. Rahmani, C. Jagadish, D. N. Neshev, and M. Kauranen, "Resonant harmonic generation in AlGaAs nanoantennas probed by cylindrical vector beams," Nanoscale **11**, 1745–1753 (2019).
29. G. Milione, H. I. Sztul, D. A. Nolan, and R. R. Alfano, "Higher-order Poincaré sphere, Stokes parameters, and the angular momentum of light," Phys. Rev. Lett. **107**, 053601 (2011).
30. Z. Bomzon, G. Biener, V. Kleiner, and E. Hasman, "Radially and azimuthally polarized beams generated by space-variant dielectric subwavelength gratings," Opt. Lett. **27**, 285–287 (2002).
31. C. Maurer, A. Jesacher, S. Fürhapter, S. Bernet, and M. Ritsch-Marte, "Tailoring of arbitrary optical vector beams," New J. Phys. **9**, 78–78 (2007).
32. E. Karimi, B. Piccirillo, E. Nagali, L. Marrucci, and E. Santamato, "Efficient generation and sorting of orbital angular momentum eigenmodes of light by thermally tuned q-plates," Appl. Phys. Lett. **94**, 231124 (2009).
33. H. Chen, J. Hao, B.-F. Zhang, J. Xu, J. Ding, and H.-T. Wang, "Generation of vector beam with space-variant distribution of both polarization and phase," Opt. Lett. **36**, 3179–3181 (2011).
34. D. Naidoo, F. S. Roux, A. Dudley, I. Litvin, B. Piccirillo, L. Marrucci, and A. Forbes, "Controlled generation of higher-order Poincaré sphere beams from a laser," Nat. Photonics **10**, 327–332 (2016).
35. I. A. Litvin, A. Dudley, F. S. Roux, and A. Forbes, "Azimuthal decomposition with digital holograms," Opt. Express **20**, 10996–11004 (2012).
36. F. Gori, "Polarization basis for vortex beams," J. Opt. Soc. Am. A **18**, 1612–1617 (2001).
37. R. Martínez-Herrero and P. M. Mejías, "Stokes-parameters representation in terms of the radial and azimuthal field components: A proposal," Opt. & Laser Technol. **42**, 1099–1102 (2010).
38. E. Wolf, "Electromagnetic diffraction in optical systems I. An integral representation of the image field," Proc. Royal Soc. London. Ser. A. Math. Phys. Sci. **253**, 349–357 (1959).
39. B. Richards and E. Wolf, "Electromagnetic diffraction in optical systems II. Structure of the image field in an aplanatic system," Proc. Royal Soc. Lond. A: Math. Phys. Eng. Sci. **253**, 358–379 (1959).
40. S. Pereira and A. van de Nes, "Superresolution by means of polarisation, phase and amplitude pupil masks," Opt. Commun. **234**, 119–124 (2004).
41. H. Wang, L. Shi, B. Lukyanchuk, C. Sheppard, and C. T. Chong, "Creation of a needle of longitudinally polarized light in vacuum using binary optics," Nat. Photonics **2**, 501–505 (2008).
42. L. Turquet, J.-P. Kakko, X. Zang, L. Naskali, L. Karvonen, H. Jiang, T. Huhtio, E. Kauppinen, H. Lipsanen, M. Kauranen, and G. Bautista, "Tailorable second-harmonic generation from an individual nanowire using spatially phase-shaped beams," Laser Photonics Rev. **11**, 1600175 (2017).
43. L. W. Casperson, "Gaussian light beams in inhomogeneous media," Appl. Opt. **12**, 2434–2441 (1973).
44. T. Bauer, P. Banzer, E. Karimi, S. Orlov, A. Rubano, L. Marrucci, E. Santamato, R. W. Boyd, and G. Leuchs, "Observation of optical polarization Möbius strips," Science **347**, 964–966 (2015).
45. J. Mäkitalo, S. Suuriniemi, and M. Kauranen, "Boundary element method for surface nonlinear optics of nanoparticles," Opt. Express **19**, 23386–23399 (2011).
46. Jouni Mäkitalo, "Boundary Integral Operators in Linear and Second-order Nonlinear Nano-optics," Dissertations, Tampere University of Technology, Tampere (2015).
47. P. B. Johnson and R. W. Christy, "Optical constants of the noble metals," Phys. Rev. B **6**, 4370–4379 (1972).